\documentclass[prd,twocolumn,showpacs,nofootinbib,preprintnumbers,floatfix]{revtex4}  % for review and submission
\usepackage[plainpages=false, colorlinks=true, anchorcolor=blue, linkcolor=blue, citecolor=blue, bookmarks=false]{hyperref}
\newcommand{\bqn}{\begin{eqnarray}}
\newcommand{\eqn}{\end{eqnarray}}
\usepackage{graphicx}  % needed for figures
\usepackage{subcaption}
\usepackage{dcolumn}   % needed for some tables
\usepackage{bm}        % for math
\usepackage{amssymb}   % for math
\usepackage{natbib}
\usepackage{amsmath}
\usepackage{float}
\usepackage[utf8]{inputenc}
\usepackage{booktabs}

\begin{document}

\newcommand{\apjl}{Astrophys. J. Lett.}
\newcommand{\apjs}{Astrophys. J. Suppl. Ser.}
\newcommand{\aap}{Astron. \& Astrophys.}
\newcommand{\aj}{Astron. J.}
\newcommand{\pasp}{PASP}
\newcommand{\araa}{Ann. Rev. Astron. Astrophys. }
\newcommand{\aapr}{Astronomy and Astrophysics Review}
\newcommand{\ssr}{Space Science Reviews}
\newcommand{\mnras}{Mon. Not. R. Astron. Soc.}
\newcommand{\apss}{Astrophys. and Space Science}
\newcommand{\jcap}{JCAP}
\newcommand{\na}{New Astronomy}
\newcommand{\pasj}{PASJ}
\newcommand{\pasa}{Pub. Astro. Soc. Aust.}
\newcommand{\physrep}{Physics Reports}

\title{Revisiting the Constancy of the Speed of Light: Galaxy Cluster Mass Bias Implications}

\author{R. F. L. Holanda$^{1,2}$}\email{holandarfl@fisica.ufrn.br}
\author{Marcelo Ferreira$^{1}$}\email{fsm.fisica@gmail.com}
\author{Javier E. Gonzalez$^3$} \email{javiergonzalezs@academico.ufs.br}
\author{S. H. Pereira$^{3}$}\email{s.pereira@unesp.br}

\affiliation{$^1$Departamento de F\'{\i}sica, Universidade Federal do Rio Grande do Norte, Natal - Rio Grande do Norte, 59072-970, Brazil.}
\affiliation{$^2$Departamento de F\'{\i}sica, Universidade Federal de Campina Grande, 58429-900, Campina Grande - PB, Brazil}
\affiliation{$^3$Departamento de F\'{\i}sica,
Faculdade de Engenharia e Ciências de Guaratinguetá,
Universidade Estadual Paulista (UNESP), Av. Dr. Ariberto Pereira da Cunha 333, 12516-410, Guaratinguetá, SP, Brazil.}
\affiliation{$^4$Departamento de F\'{i}sica, Universidade Federal de Sergipe, São Cristóvão, SE 49107-230, Brazil.}

\begin{abstract}

In recent years, improvements in galaxy cluster observations have enabled a variety of tests of fundamental physics using these systems. In this work, we test the constancy of the speed of light, $c$, by combining X-ray gas mass fraction measurements from galaxy clusters with SNe Ia luminosity distance measurements from Pantheon+. We adopt the SH0ES prior on $H_0$ and the $\Omega_b/\Omega_m$ ratio from galaxy clustering observations, thereby minimizing the dependence of our analysis on any specific cosmological model. We explore different assumptions for the cluster mass calibration (mass bias), including \textsc{CLASH}, \textsc{CCCP}, and Planck-based estimates. We find no deviation from a constant $c$ when adopting \textsc{CLASH} or \textsc{CCCP} priors, while Planck-based calibration yields a mild tension, with the hypothesis of constant $c$ being only marginally consistent at the $2\sigma$ level, indicating a non-negligible sensitivity of the results to the adopted calibration scheme.

\end{abstract}


\maketitle

\section{Introduction}

The standard model of physics assumes that some quantities of nature are true constants along the evolution of the universe, such as the Newtonian constant ($G$), the fine structure constant ($\alpha$), the Planck constant ($h$), the Boltzmann constant ($k_B$), and the speed of light ($c$) (see, e.g., \cite{Uzan:2010pm,2017RPPh...80l6902M}). Testing the constancy of these quantities is of fundamental importance, as any detected variation would have profound implications for both fundamental physics and cosmology. 
In the electromagnetic sector, several methods have been developed to probe possible variations of the fine structure constant $\alpha$, including observations of high-redshift quasar absorption systems \cite{Dzuba:1998au,Webb:1998cq,King:2012id,Kotus:2016xxb}. Similarly, different approaches have been proposed to test the constancy of the speed of light, combining cosmological probes such as SNe Ia, BAO, $H(z)$ measurements, and strong gravitational lensing \cite{Qi:2014zja,Salzano:2016hce,Liu:2021eit,Wang:2019tdn,Rodrigues:2021wyk}. Within current observational uncertainties, these analyses have generally found no significant evidence for deviations from a constant $c$.

Galaxy clusters provide an additional and complementary avenue for testing fundamental physics. In particular, the gas mass fraction has been widely used as a cosmological probe and, more recently, as a tool to test the stability of fundamental constants \cite{Mendonca:2021eux}.  Different scenarios were explored, including both a constant and a redshift-dependent depletion factor, while the impact of different $H_0$ estimates was also examined. In most cases, no variation in the speed of light was found at a confidence level greater than $1\sigma$. However, these analyses rely on assumptions about cluster physics, especially regarding mass calibration. 
Recent studies have highlighted a tension in cluster mass estimates derived from different methods, including X-ray, gravitational lensing, and CMB observations \cite{2015ApJ...806..247B,2018A&A...610L...4H,2019MNRAS.489..401Z,2023A&A...674A..48W}. This discrepancy is commonly associated with the hydrostatic mass bias and other systematic effects related to cluster astrophysics. Since the gas mass fraction depends directly on the total mass estimate, uncertainties in mass calibration can significantly impact cosmological inferences.

In this work, we test the constancy of the speed of light by combining galaxy cluster observations with {SNe Ia data from the Pantheon+ sample \cite{Brout:2022vxf}, adopting the SH0ES estimate of the absolute magnitude $M_b$}, as well as the $\Omega_b/\Omega_m$ ratio from galaxy clustering observations \cite{2025PhRvD.111f3526K}, making our analysis only weakly dependent on the background cosmology.
The key novelty of our analysis is a detailed investigation of how different assumptions on the \textit{mass bias} affect constraints on a possible variation in the speed of light. { Galaxy clusters are promising systems to probe variations of fundamental constants: their X-ray and SZ observables depend on the physical properties of the intracluster medium and on assumptions about its thermodynamic and dynamical state, while their total masses can be independently calibrated through gravitational lensing. Nevertheless, the known tension among mass-bias estimates derived from X-ray, lensing, and CMB analyses must be properly quantified, since the mass calibration directly controls the robustness of constraints on $\Delta c/c$ and on derived cosmological parameters.}
In particular, we assess the sensitivity of our results to three distinct Gaussian priors for the mass bias parameter: one derived from the CLASH sample~\cite{sereno2015comparing}, another from the Canadian Cluster Comparison Project (CCCP)~\cite{herbonnet2020cccp}, and a third based on a joint analysis combining Planck primary CMB data, Planck-SZ number counts, the Planck thermal SZ power spectrum, and BAO measurements~\cite{salvati2018constraints}. We consider two phenomenological parameterizations for a redshift-dependent speed of light, \(c(z) = c_0 \left(1 + c_1 z\right)\) and \(c(z) = c_0 \left(1 + \frac{c_1 z}{1+z}\right)\), where \(c_0\) denotes the present-day value. By combining multiple datasets and explicitly accounting for uncertainties in cluster mass calibration, our analysis provides a robust framework to assess the impact of astrophysical systematics on tests of the constancy of the speed of light. This systematic treatment of mass bias represents a significant step forward compared to previous analyses, where this source of uncertainty was often fixed or neglected. The manuscript is organized as follows. In Sec.~\ref{methodology} we present the methodology. The data are described in Sec.~\ref{data}. The analysis and results are presented in Sec.~\ref{sec:analysis}, and the conclusions are given in Sec.~\ref{sec:conclusions}.

\section{Methodology}
\label{methodology}

\begin{figure*}[htb]
    \centering
    \begin{subfigure}{0.45\textwidth}
        \centering
        \includegraphics[width=\linewidth]{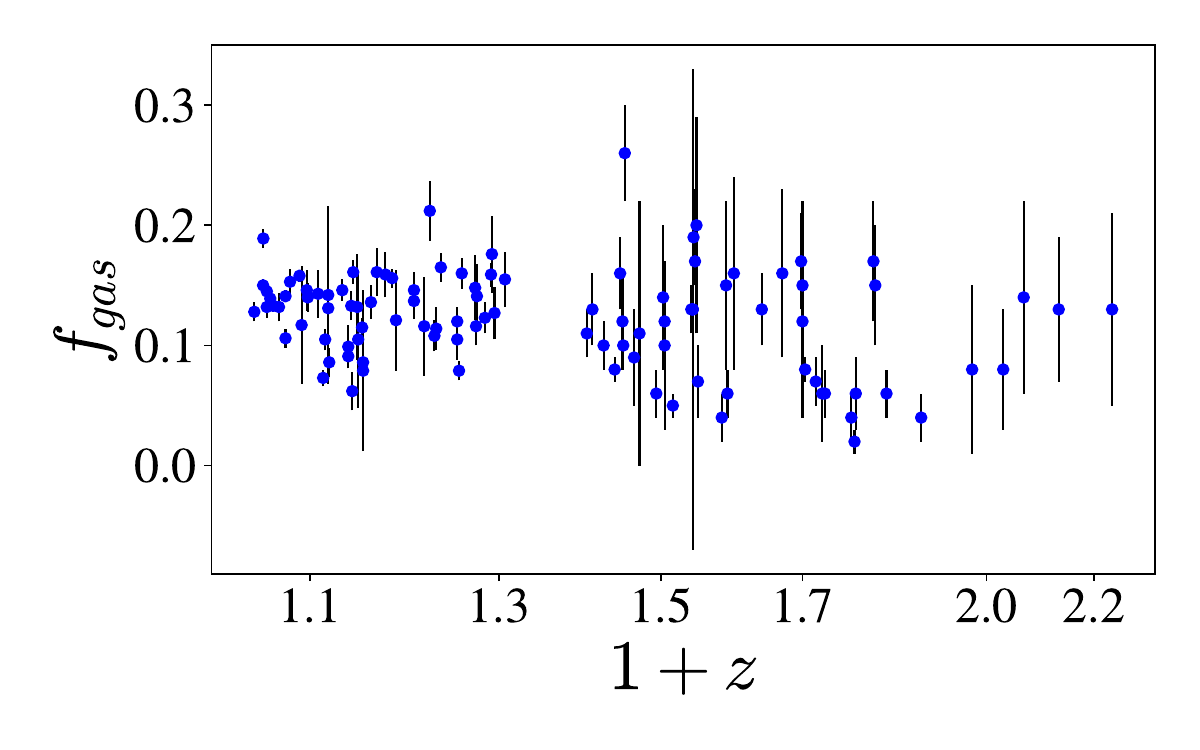}
    \end{subfigure}
    \hfill
    \begin{subfigure}{0.45\textwidth}
        \centering
        \includegraphics[width=\linewidth]{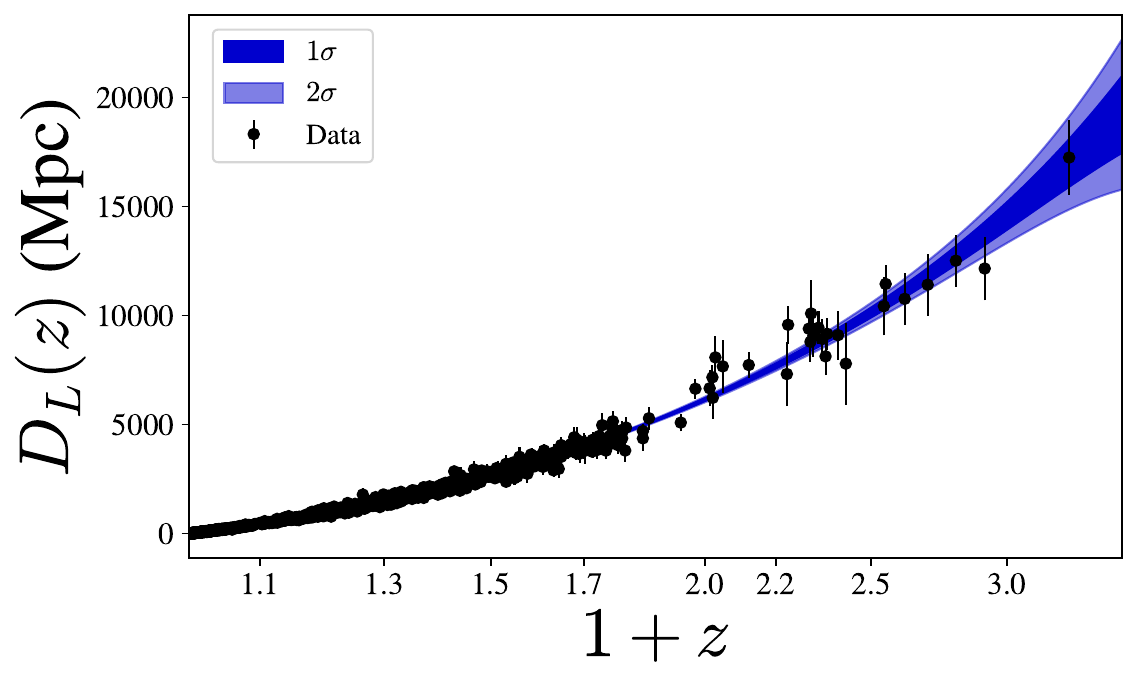}
    \end{subfigure}
    \caption{Left: Gas mass fraction measurements used in this work. Right: {Gaussian Process reconstruction of $D_L(z)$ based on the Pantheon+ data.} The shaded regions indicate the $1\sigma$ and $2\sigma$ confidence intervals, and the black points show the observational data. %{\bf The horizontal axis shows $1+z$ in logarithmic scale.}
    }
    \label{bao_fgas}
\end{figure*}

The baryonic matter content of galaxy clusters is mostly composed of intracluster gas. This diffuse gas primarily appears in the X-ray spectrum through the thermal bremsstrahlung process. In this context, a cosmologically significant quantity is the gas mass fraction\footnote{This quantity has been widely used as a cosmological probe in different contexts (e.g., \cite{Allen2007, Allen2011, Mantz:2014xba}), including tests of the stability of fundamental constants \cite{Mendonca:2021eux, Galli:2012bf}.}, which is defined as the ratio between the mass of the intracluster gas and the total mass (including dark matter):
\begin{equation}
    \label{mass_frac}
    f_{gas} \equiv \frac{M_{gas}}{M_{tot}}.
\end{equation}

The total mass within a given radius $R$ can be obtained assuming hydrostatic equilibrium \cite{2011ARA&A..49..409A}:
\begin{equation}
    M_{tot}(<R) = - \frac{k_{\mathrm{B}} T R}{G \mu m_{\mathrm{H}}} \frac{d \ln n_{\mathrm{e}}(r)}{d \ln r} \bigg|_{r=R},
\end{equation}
where $T$ and $n_{\mathrm{e}}$ are the electronic temperature and density, respectively, while $m_{\mathrm{H}}$ and $\mu$ are the mass and mean molecular weight of hydrogen, respectively. On the other hand, the gas mass obtained from X-ray observations can be written as \cite{sarazin1986x}:
\begin{eqnarray}
    \label{mgas}
    M_{gas}(<R) = \left( \frac{3 \pi \hbar m_{\mathrm{e}} c^2}{2(1+X) e^2} \right)^{1/2} \left( \frac{3 m_{\mathrm{e}} c^2}{2 \pi k_{\mathrm{B}} T} \right)^{1/4} m_{\mathrm{H}} \nonumber \\
    \times \frac{r_c^{3/2}}{g_{\mathrm{B}}(T)^{1/2}} \left[ \frac{I_{\mathrm{M}}(R/r_{\mathrm{c}}, \beta)}{I_{\mathrm{L}}^{1/2}(R/r_{\mathrm{c}}, \beta)} \right] [L_{\mathrm{x}}(<R)]^{1/2},
\end{eqnarray}
where $X \approx 0.76$ represents the hydrogen mass fraction, $m_{\mathrm{e}}$ is the electron mass, $r_c$ is the core radius, $g_{\mathrm{B}}(T)$ is the Gaunt factor, and
\begin{equation*}
    I_{\mathrm{M}}(y, \beta) \equiv \int_0^y (1 + x^2)^{-3 \beta/2} x^2 dx,
\end{equation*}
\begin{equation*}
    I_{\mathrm{L}}(y, \beta) \equiv \int_0^y (1 + x^2)^{-3 \beta} x^2 dx.
\end{equation*}
From Eq.~\eqref{mgas}, it follows that \(M_{\text{gas}}(<R) \propto c^{3/2}\). Consequently, if \(c\) is allowed to vary as \(c(z) = c_0 \phi(z)\), Eq.~\eqref{mass_frac} must be updated accordingly to
\begin{equation}
    \label{fgas1}
    f_{\mathrm{gas}} \equiv \phi(z)^{3/2} \frac{M_{\mathrm{gas}}}{M_{\mathrm{tot}}}.
\end{equation}

Now, as a cosmological tool, the X-ray gas mass fraction can be modeled using the following expression \cite{eckert2019non}:
\begin{equation}
    \label{fgas2}
    f_{gas} = K \gamma(z) \frac{\Omega_{\mathrm{b}}}{\Omega_{\mathrm{m}}} \left(\frac{D_L^*}{D_L} \right)^{3/2} - f_{\star},
\end{equation}
where $\gamma(z)$ is the baryon depletion factor and $f_{\star}$ represents the stellar fraction, indicating the proportion of baryons confined in stars. %{\bf From Eq.~\eqref{mgas}, one finds that the gas mass fraction scales as $f_{\rm gas} \propto L(<R)^{1/2}$. On the other hand, the quantities $r_c$ and $R$ correspond to physical sizes of the cluster and therefore scale with the angular diameter distance as $r_c \propto D_A(z)$ and $R \propto D_A(z)$. Since the X-ray luminosity scales as $L \propto D_L^2(z)$, the combined dependence leads to $f_{\rm gas} \propto D_L^{3/2}(z)$, assuming the validity of the cosmic distance duality relation (CDDR). This is the origin of the $D_L^{3/2}$ scaling appearing in Eq.~\eqref{fgas2} (see, e.g., \cite{sarazin1986x, Allen2007, Ettori2009}).}
%From Eq.~\eqref{mgas}, one finds that $M_{\rm gas} \propto L_X^{1/2}$. However, t
{ The origin of the $D_L^{3/2}$ scaling can be understood from the dependence of X-ray observables on cluster properties and cosmological distances.
The X-ray surface brightness is proportional to the line-of-sight integral of the square of the electron density, $S_X \propto \int n_e^2 dV$, while the observed flux scales as $S_X \propto L_X / D_L^2$. Therefore, the electron density scales as $n_e \propto (L_X / D_L^2)^{1/2}$. 
The gas mass is obtained from $M_{\rm gas} \propto n_e V$, where the emitting volume scales as $V \propto D_A^3$, since cluster sizes are inferred from angular measurements. Combining these relations, one finds
\begin{equation*}
M_{\rm gas} \propto L_X^{1/2} D_A^{3/2}.
\end{equation*}
On the other hand, the total mass inferred from hydrostatic equilibrium scales as $M_{\rm tot} \propto R \propto D_A$. Therefore, the gas mass fraction scales as
\begin{equation*}
f_{\rm gas} \propto \frac{M_{\rm gas}}{M_{\rm tot}} \propto L_X^{1/2} D_A^{1/2}.
\end{equation*}
%The observed flux is related to the intrinsic luminosity via $L_X = 4\pi D_L^2 f_X$.
Finally, using the cosmic distance duality relation $D_L = (1+z)^2 D_A$, one obtains
\begin{equation*}
f_{\rm gas} \propto D_L^{3/2}.
\end{equation*}
 This is the origin of the $D_L^{3/2}$ scaling appearing in Eq.~\eqref{fgas2}. This scaling is well established in the literature  \cite{sarazin1986x, Allen2007, Ettori2009}.}

We follow Ref.~\cite{corasaniti2021cosmological}, where the parameter $K$ accounts for the mass calibration bias and is defined as $K \equiv 1 - b = M^{HE} / M^{tot}$, with $M^{HE}$ being the cluster mass estimated under the assumption of hydrostatic equilibrium. Here, $D_L^*$ represents the luminosity distance of the fiducial cosmological model, typically a flat $\Lambda$CDM model with $\Omega_m=0.3$ and $H_0=70$ km/s/Mpc, adopted to infer the gas mass fraction measurements.  In our analysis, we assume an additional 10\% uncertainty associated with instrument calibration, X-ray modeling, and non-thermal pressure support, which must be properly accounted for when deriving cosmological constraints, as discussed by \cite{2008MNRAS.383..879A}.

Considering the relation in Eq.~\eqref{fgas1}, the above equation should be corrected as follows:
\begin{equation}
    \label{fgasphi}
    f_{gas} = \phi(z)^{3/2} K \gamma(z) \frac{\Omega_{\mathrm{b}}}{\Omega_{\mathrm{m}}} \left(\frac{D_L^*}{D_L} \right)^{3/2} - f_{\star},
\end{equation}
which can be used to obtain limits on $\phi(z)$ if one knows the luminosity distance to a galaxy cluster. In this work, we consider $\phi(z) = (1 + c_1 z)$ ($P_1$) and $\phi(z) = [1 + c_1 z / (1+z)]$ ($P_2$), where $c_1$ is a free parameter.

In the following analysis, we adopt the $\Omega_b/\Omega_m$ ratio derived from recent galaxy clustering observations \cite{2025PhRvD.111f3526K}, namely $\Omega_b/\Omega_m = 0.173 \pm 0.027$. The choice of using the SH0ES $M_{\mathrm{b}}$ prior and the $\Omega_{\mathrm{b}}/\Omega_{\mathrm{m}}$ ratio from galaxy clustering aims to ensure that our analysis remains as model-independent as possible, allowing us to test the constancy of the speed of light without assuming a specific cosmological background.  The baryon depletion factor $\gamma(z)$ is treated here as an astrophysical systematic entering the $f_{\rm gas}$ modeling. We adopt the redshift dependence $\gamma^{\text{FABLE}}_{500}(z)=0.931(1+0.017z+0.003z^2)$ as a physically motivated baseline calibrated by hydrodynamical simulations \cite{10.1093/mnras/staa2235,corasaniti2021cosmological}. The intrinsic scatter $\sigma_{\gamma}=0.04$ represents the simulation-calibrated cluster-to-cluster dispersion of the depletion factor at fixed redshift and mass scale, and it is included to account for residual uncertainties related to baryonic physics not captured by the mean trend. In practice, the uncertainty in $\gamma(z)$, including this intrinsic scatter, is propagated into the final inference through the covariance $C_\xi$. As described in Sec.~\ref{sec:analysis}, we obtain $\xi$ and $C_\xi$ via Monte Carlo sampling of all quantities entering Eq.~\eqref{fgas2}, including $\gamma$, according to their quoted uncertainties and adopted priors. This procedure consistently transfers the scatter of $\gamma(z)$ into the likelihood without introducing $\gamma$ as an additional sampled parameter in the MCMC. Moreover, by considering Eq.~(8) from Ref.~\cite{2022A&A...663L...6A}, we multiply this factor ($\gamma$) by $w \beta x^{\gamma} + \delta w$, where $w = M_{500}/(5 \times 10^{14} \, h^{-1} M_{\odot})$, $x = r/R_{500,c}$, while $\beta$, $\gamma$, and $\delta$ are the free parameters: $(\beta, \gamma, \delta) = (0.12, 0.23, -0.22)$. This takes into account a possible radial and mass dependence of $\gamma(z)$.

For the stellar fraction, we use a Gaussian prior of $f_{\star,500} = 0.015 \pm 0.005$, aligned with estimates obtained from cluster samples with masses comparable to those in the gas mass fraction dataset used in this work \cite{eckert2019non}.  The stellar mass fraction is known to exhibit some mass dependence \cite{Chiu_2018}. However, in the massive-cluster regime relevant for this work, different observational methods converge to a consistent mean value with limited scatter \cite{Leauthaud_2011, Ettori2009, eckert2019non, eckert2016xxl, corasaniti2021cosmological}. Moreover, the stellar component represents a subdominant contribution compared to the gas mass, so uncertainties in $f_{\star,500}$ have a negligible impact on our results.

To ensure a conservative approach, we tested our results under varying assumptions for the mass calibration bias $K$ found in the literature. We consider the following Gaussian priors:
\begin{itemize}
    \item $K^{\text{CCCP}}_{500} = 0.84 \pm 0.04$, derived from the analysis of a sample of clusters as part of the Canadian Cluster Comparison Project (CCCP) \cite{herbonnet2020cccp}.
    \item $K^{\text{CLASH}}_{500} = 0.78 \pm 0.09$, consistent with estimates from the analysis of the CLASH (Cluster Lensing And Supernova survey with Hubble) sample \cite{sereno2015comparing}.
    \item $K^{\text{CMB}}_{500} = 0.65 \pm 0.04$, consistent with the mass bias inferred from the joint analysis of the Planck primary CMB, Planck-SZ number counts, Planck thermal SZ power spectrum, and BAO \cite{salvati2018constraints}.
\end{itemize}

A key aspect of our analysis is that the mass calibration parameter is treated as a dominant systematic in the $f_{\rm gas}$ modeling. Therefore, the Gaussian priors $K_{500}^{\rm CCCP}$, $K_{500}^{\rm CLASH}$, and $K_{500}^{\rm CMB}$ are not assumed to represent a baseline truth for the X-COP subsample (see Sec.~\ref{data}). Instead, we perform the inference separately under each prior and compare the resulting constraints on $c_1$, thereby quantifying the sensitivity of $\Delta c/c$ to the current spread of mass-bias determinations in the literature. In this sense, our approach constitutes a robustness analysis against mass-calibration systematics and does not rely on the assumption that X-COP shares the same intrinsic mass bias as CLASH or CCCP, which have different selection functions and baryonic properties. 

{ The $K_{\rm CLASH}$ calibration is based on lensing measurements of a relatively small sample of massive clusters and may be affected by residual systematics in the comparison between X-ray and lensing masses. The $K_{\rm CMB}$ calibration, on the other hand, is derived from a global analysis combining Sunyaev--Zel’dovich cluster counts and the CMB power spectrum, and is therefore sensitive to degeneracies between cosmological parameters and cluster physics. Since the gas mass fraction in the 103-cluster sample is defined as an integrated quantity within $r_{500}$, it is also sensitive to baryonic effects, including the stellar component and the thermodynamical state of the intracluster medium, which may contribute to small shifts in the inferred quantities. We also note that the apparent offset at low redshift ($z \lesssim 0.3$), particularly for the $K_{\rm CLASH}$ and $K_{\rm CMB}$ calibrations, can be understood as a consequence of the different systematics underlying these mass-bias estimates. }

\section{{ Data}}
\label{data}

Here, we outline the datasets used in the methodology described in the previous section:

\begin{itemize}
    \item \textbf{Gas mass fraction:} In this work, we used measurements of the gas mass fraction within $R_{500}$, { where $R_{500}$ is defined as the radius within which the mean enclosed density is 500 times the critical density at the cluster redshift.} Our dataset consists of the gas mass fractions of 12 clusters at $z \lesssim 0.1$ from X-COP, analyzed in \cite{eckert2019non}; 44 clusters in the range $0.1 \lesssim z \lesssim 0.3$ investigated in \cite{ettori2010mass}; and 47 clusters at $0.4 \lesssim z \lesssim 1.2$ studied in \cite{ghirardini2017evolution}. These 103 data points are plotted in Fig.~\ref{bao_fgas}. {These clusters are massive systems with $M_{500} \geq 10^{14} M_\odot$,} and the fiducial cosmology assumed in all these analyses is a flat $\Lambda$CDM model with $\Omega_m = 0.3$ and $h = 0.7$.

\begin{figure*}[htb]
    \centering
    \begin{subfigure}{0.3\textwidth}
        \centering
        \includegraphics[width=\linewidth]{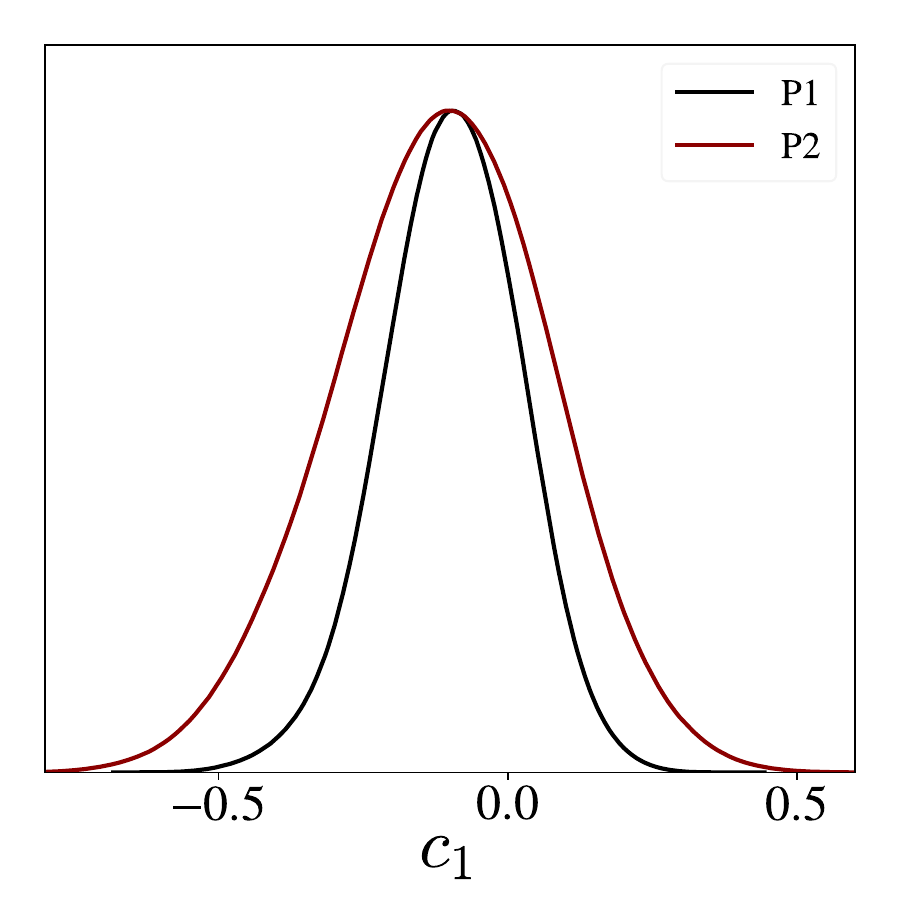}
    \end{subfigure}
    \hfill
    \begin{subfigure}{0.3\textwidth}
        \centering
        \includegraphics[width=\linewidth]{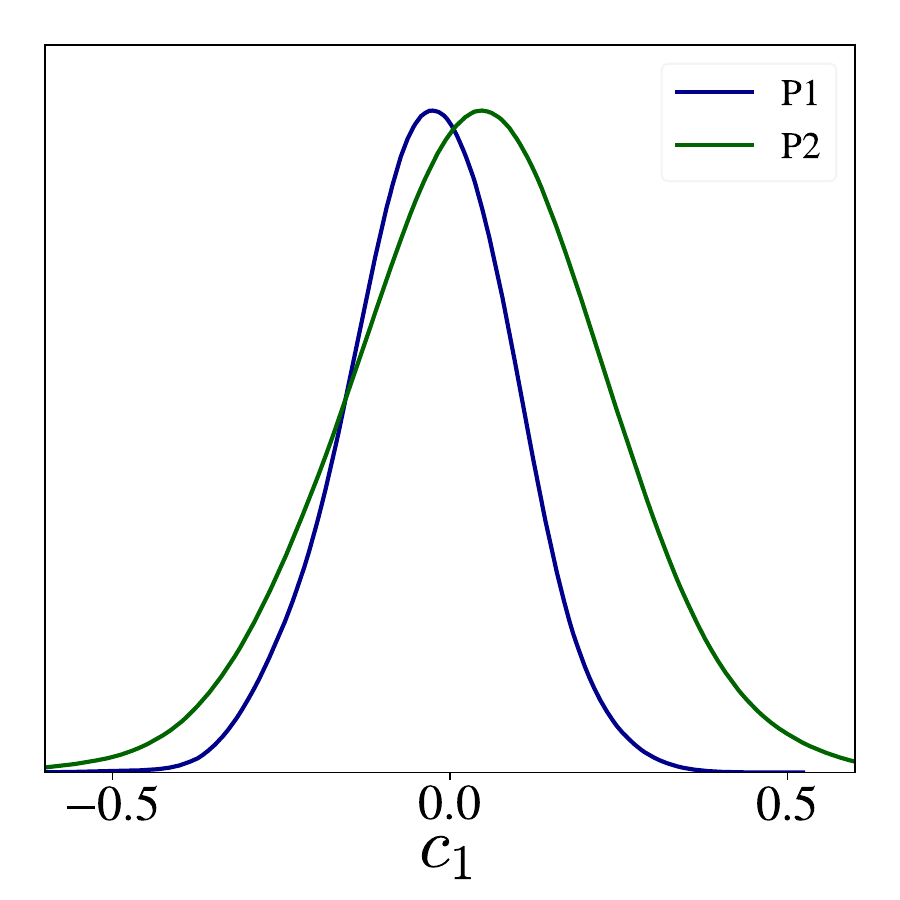}
    \end{subfigure}
    \hfill
    \begin{subfigure}{0.3\textwidth}
        \centering
        \includegraphics[width=\linewidth]{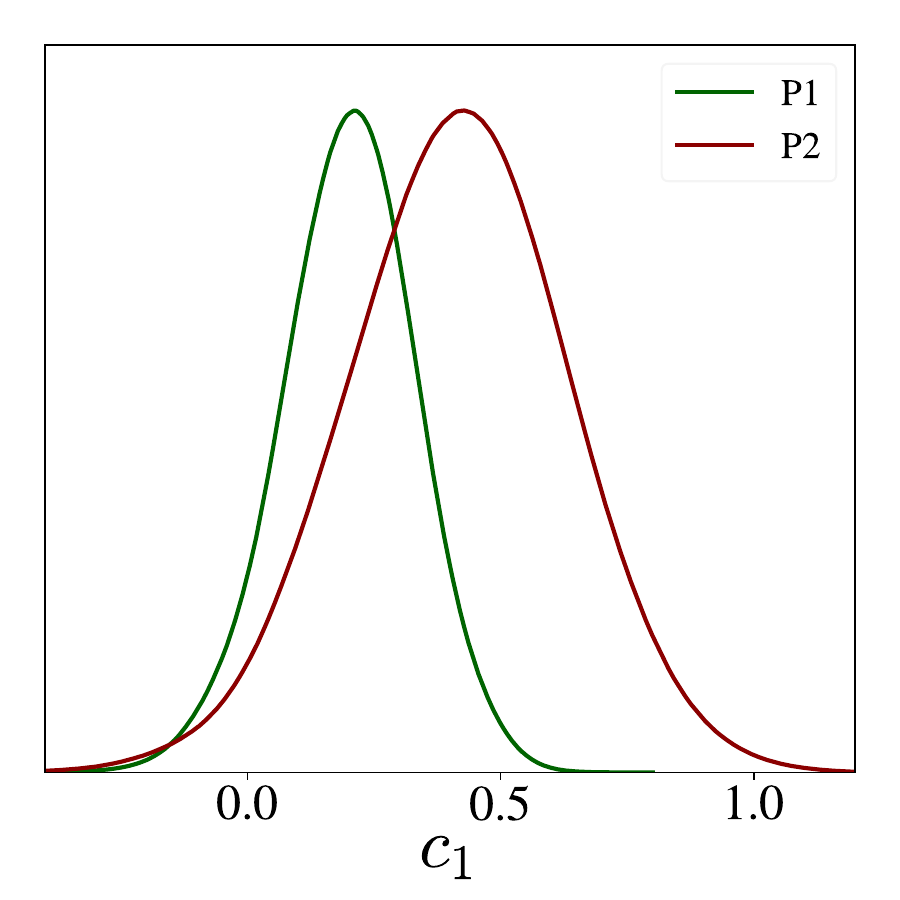}
    \end{subfigure}
    \caption{Left: Posterior probability distribution of $c_1$ with {$K_{500}^{CCCP}$}. Center: Posterior probability distribution of $c_1$ with $K_{500}^{CLASH}$. Right: Posterior probability distribution of $c_1$ with $K_{500}^{CMB}$.}
    \label{fig:results1}
\end{figure*}

\begin{table*}[htbp]
    \centering
    \renewcommand{\arraystretch}{1.8}
    \setlength{\tabcolsep}{18pt}
    \begin{tabular}{lcc}
        \hline
        \textbf{Astrophysics prior} & $c_1(P_1)$ & $c_1(P_2)$ \\
        \hline
        \hline
        $K_{500}^{CCCP} + \gamma_{500}^{FABLE}$ & $-0.10 \pm 0.12$ & $-0.11 \pm 0.19$ \\
        $K_{500}^{CLASH} + \gamma_{500}^{FABLE} $ & $-0.03 \pm 0.12$ & $0.04 \pm 0.19$ \\
        $K_{500}^{CMB} + \gamma_{500}^{FABLE} $ & $0.20 \pm 0.13$ & $0.41 \pm 0.22$ \\
        $K_{500}^{CCCP} + \gamma_{500}^{The300}$ & $-0.09 \pm 0.11$ & $-0.11 \pm 0.19$ \\
        $K_{500}^{CLASH} + \gamma_{500}^{The300} $ & $-0.01 \pm 0.12$ & $0.04 \pm 0.19$ \\
        $K_{500}^{CMB} + \gamma_{500}^{The300} $ & $0.24 \pm 0.13$ & $0.47 \pm 0.21$ \\
        \hline
    \end{tabular}
    \caption{Summary of the constraints on the parameter $c_1$ (at $1\sigma$) obtained using different mass-bias and baryon depletion priors.}
    \label{tab:results}
\end{table*}

    \item \textbf{Luminosity distance from SNe Ia:}

{We used a dataset of apparent magnitudes $m_{\mathrm{b}}$ from the Pantheon+ sample \cite{Brout:2022vxf}, which consists of 1,701 light curves from 1,550 spectroscopically confirmed Type Ia supernovae in the redshift range $0.001 < z < 2.26$ (Fig.~\ref{bao_fgas}). The apparent-magnitude dataset is transformed into luminosity distances for each supernova by using the relation}
\begin{equation}
D_L = 10^{(m_{\mathrm{b}} - M_{\mathrm{b}} - 25)/5} \ \mathrm{Mpc},
\end{equation}
{where $M_b$ denotes the absolute magnitude. In this study, we assume \( M_b = -19.253 \pm 0.027 \), corresponding to the absolute magnitude compatible with the \( H_0 \) value inferred by SH0ES \cite{riess2022}.}

{The Pantheon+ dataset also includes the covariance matrix of $m_{\mathrm{b}}$ for each supernova, which can be transformed into a covariance matrix for $D_{\mathrm{L}}$ through the relation}
\begin{eqnarray}
\label{cov}
\text{cov}({\bf D_L, D_L}) = \left(  \frac{\partial {\bf D_L}}{\partial {\bf m_b}} \right) \text{cov}({\bf m_b, m_b}) \nonumber \\
         \times \left(  \frac{\partial {\bf D_L}}{\partial {\bf m_{\mathrm{b}}}} \right)^T,
\end{eqnarray}
{where the variables in bold correspond to vector representations of each dataset, and $\left(  \frac{\partial {\bf D_L}}{\partial {\bf m_{\mathrm{b}}}} \right)$ is the Jacobian matrix of the transformation.}

In addition, we employ the Gaussian Process (GP) method, trained on the {SNe Ia sample}, to reconstruct $D_L(z)$ at the redshifts of galaxy clusters. The GP reconstruction is performed by selecting a prior mean function and a covariance kernel, as detailed in Ref.~\cite{seikel2012reconstruction}. The covariance kernel is characterized by a set of hyperparameters that determine the properties of the reconstructed function and quantify the correlation between values of the dependent variable. In this work, to minimize reconstruction bias, we adopt a zero prior mean function. For the covariance kernel, we use the standard Gaussian kernel,
\begin{equation}
    \label{gaussian_kernel}
     k(z,z') = \sigma^2 \exp \left(  - \frac{(z - z')^2}{2l^2}
     \right),
\end{equation}
where $\sigma$ and $l$ are hyperparameters that regulate the amplitude of variations in the reconstructed function and its characteristic correlation length, respectively.

The hyperparameters of the GP are estimated by maximizing the logarithm of the marginal likelihood:
\begin{equation*}
\ln {\mathcal{L}}=-\frac{1}{2}
[\bm { D_{L}}]^T [\bm {{K}}(\bm z,\bm z)+\bm {{C}}]^{-1}[\bm { D_{L}}]
\end{equation*}
\begin{equation}
\label{log_likelihood}
 -\frac{1}{2}\ln |\bm {{K}}(\bm z,\bm z)+\bm {{C}}|-\frac{n}{2}\ln 2\pi,
\end{equation}
where \( \bm{z} \) denotes the vector of redshift measurements from the {Pantheon+ dataset}, while \( \bm{{K}}(\bm{z}, \bm{z}) \) represents the covariance matrix associated with the Gaussian Process modeling, \( \bm{{C}} \) corresponds to the covariance matrix of the observational data, and \( n \) indicates the total number of data points in the sample. The elements of \( \bm{{K}}(\bm{z}, \bm{z}) \) are calculated via Eq.~\eqref{gaussian_kernel}, following \([\bm{{K}}(\bm{z}, \bm{z})]_{ij} = k(z_j, z_i)\). The Gaussian Process reconstruction of the mean value and the correlations between values at different redshifts are calculated as described in Ref.~\cite{Gonzalez:2024qjs,dinda2024modelagnosticassessmentdarkenergy}.

\end{itemize}

\begin{figure*}[htb]
    \centering
    \begin{subfigure}{0.45\textwidth}
        \centering
        \includegraphics[width=\linewidth]{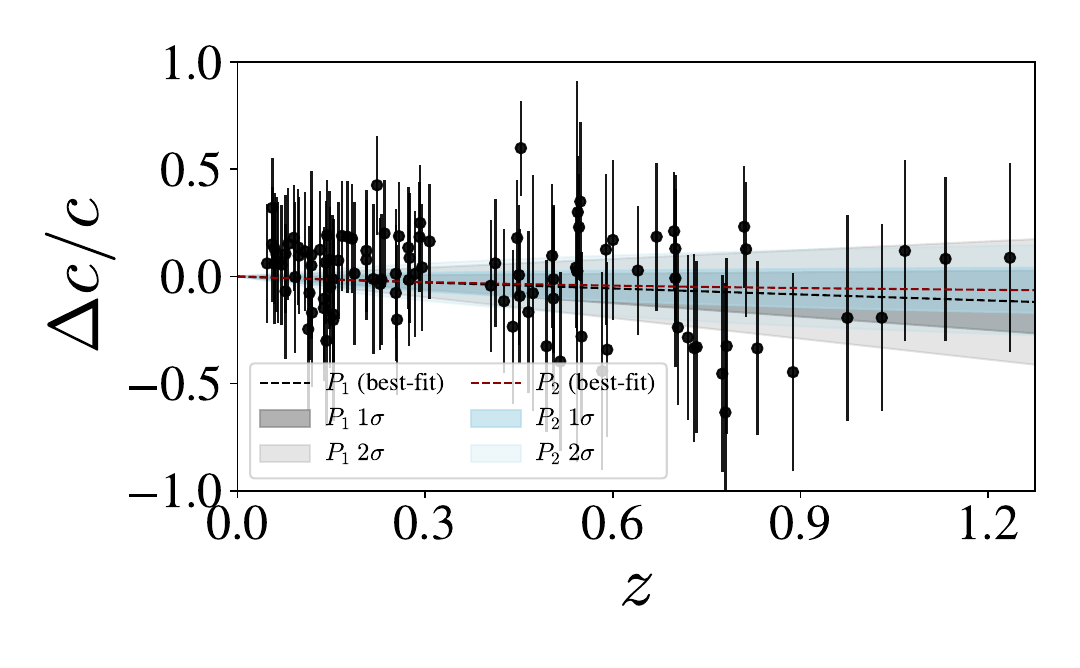}
    \end{subfigure}
    \hfill
    \begin{subfigure}{0.45\textwidth}
        \centering
        \includegraphics[width=\linewidth]{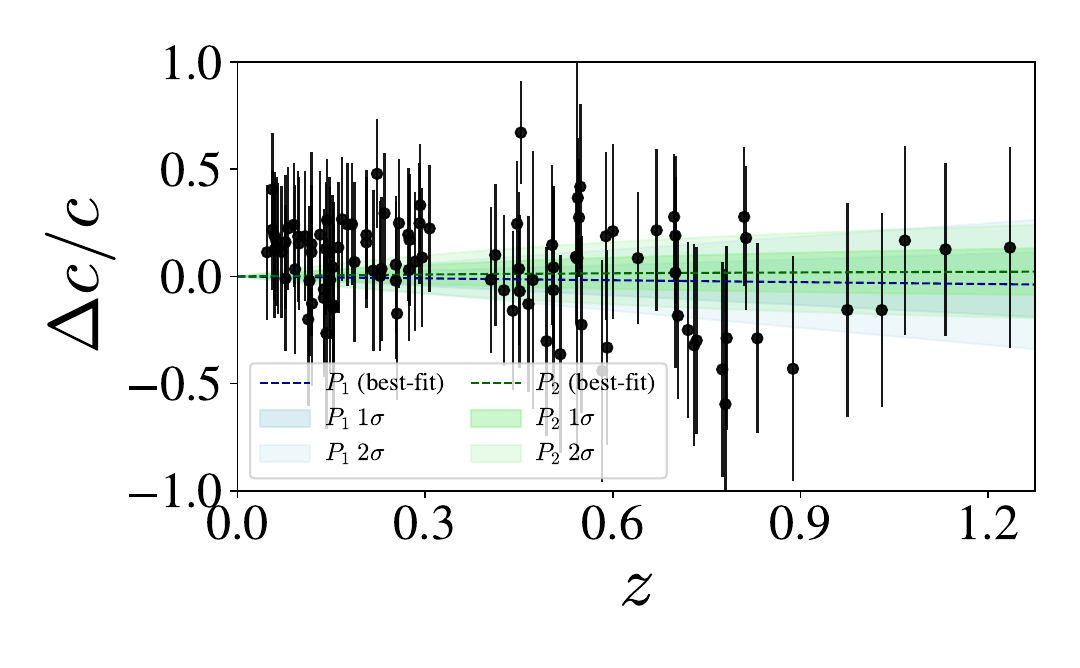}
    \end{subfigure}
    \caption{$\Delta c/c$ for different cases: {$K_{500}^{CCCP}$} (left) and $K_{500}^{CLASH}$ (right).}
    \label{fig:delta_c_cccp_clash}
\end{figure*}

\begin{figure*}[htb]
    \centering
    \begin{subfigure}{0.45\textwidth}
        \centering
        \includegraphics[width=\linewidth]{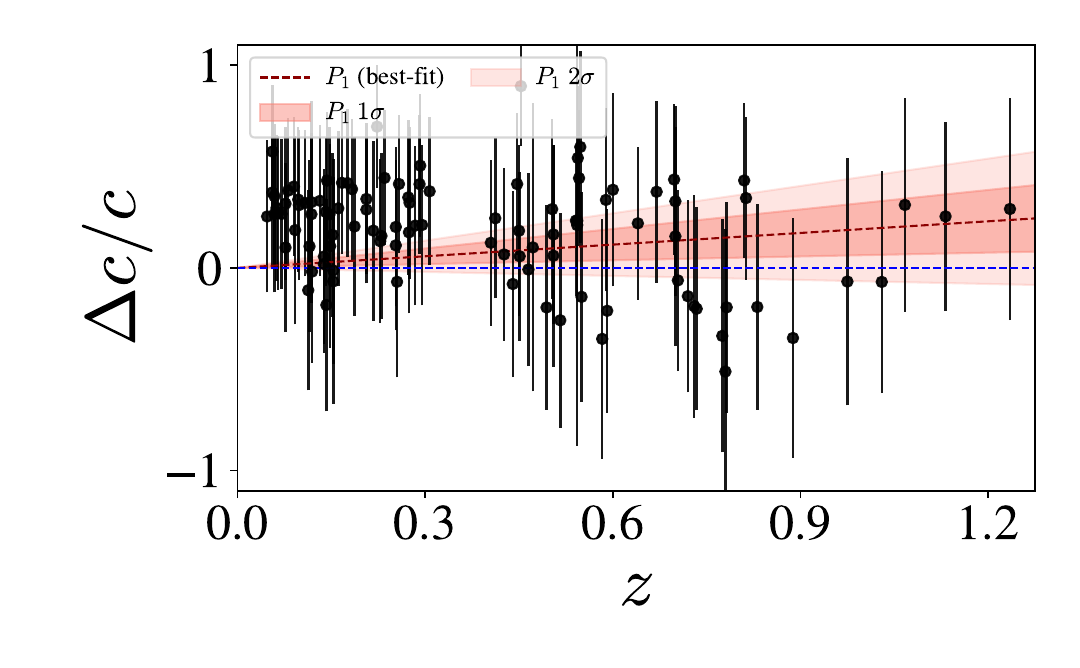}
    \end{subfigure}
    \hfill
    \begin{subfigure}{0.45\textwidth}
        \centering
        \includegraphics[width=\linewidth]{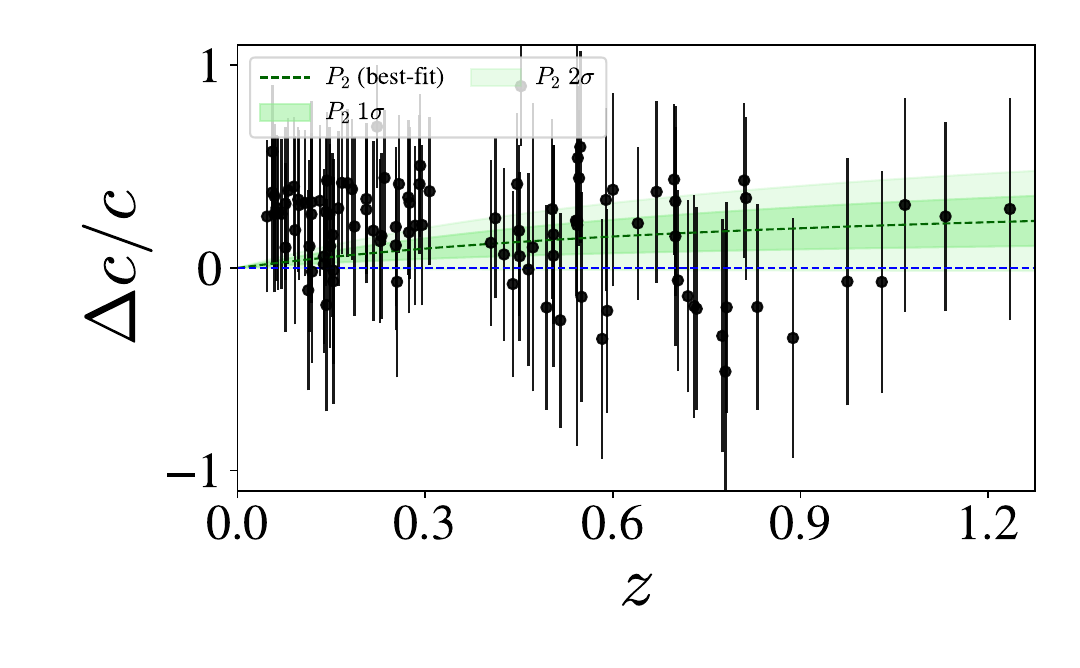}
    \end{subfigure}
    \caption{$\Delta c/c$ for {$K_{500}^{CMB}$}, considering $P_1$ (left) and $P_2$ (right).}
    \label{fig:delta_c_cmb}
\end{figure*}

{It is very important to stress that our analysis assumes the Etherington relation, $D_{\mathrm{L}}=(1+z)^2 D_{\mathrm{A}}$, in order to obtain Eq.~\eqref{fgas2}. Any variation in $c$ would also manifest itself through the standardized luminosity of SNe~Ia, but in our phenomenological framework such effects are effectively absorbed into the parameter $\Delta c/c$. Likewise, a possible violation of the  cosmic distance duality relation (CDDR)---for instance due to cosmic opacity or exotic photon interactions---would also appear as a deviation in $\Delta c/c$. However, cosmic opacity has already been extensively tested with SNe~Ia data, and so far no robust evidence for a significant bias has been found.}

\section{Analysis and Results}
\label{sec:analysis}

We utilized the Markov Chain Monte Carlo (MCMC) method to estimate the posterior probability distribution function (PDF) of the free parameter $c_1$, using the \texttt{emcee} sampler \cite{foreman2013emcee}. The constraints were obtained by sampling the likelihood function, defined as
\begin{equation}
    \mathcal{L} = \frac{1}{\sqrt{2 \pi }\vert C \vert ^{1/2}} \exp \left( - \frac{1}{2} \chi^2 \right),
\end{equation}
where the chi-squared is given by
\begin{equation}
    \chi^2 =  [ \phi(z) - \xi ] C_\xi^{-1}[ \phi(z) - \xi ]^T,
\end{equation}
with
\begin{equation}
    \xi = \left( \frac{f_{\mathrm{gas}} + f_{\star}}{K \gamma(z) (\Omega_{\mathrm{b}}/\Omega_{\mathrm{m}}) [D_L^* / D_L]^{3/2}} \right),
\end{equation}
{ where $\phi(z)$ and $\xi$ are vectors of dimension equal to the number of clusters, $C_\xi$ is the corresponding covariance matrix and it is obtained through forward Monte Carlo error propagation by considering the errors and covariance matrices of the quantities needed to estimate $\xi$, including correlations from observables such as SNe Ia, which are modeled using a multivariate Gaussian distribution with the full covariance matrix, while for the other quantities, we assume independent one-dimensional Gaussian distributions.}
We consider $\phi(z) = (1 + c_1 z)$ ($P_1$) and $\phi(z) = (1 + c_1 z/(1+z))$ ($P_2$), where $c_1$  is the only free parameter of the model. These parametrizations allow us to test both linear and nonlinear deviations from the standard constant value as a function of redshift.

{ The PDF is proportional to the product of the likelihood and the prior:
%\begin{eqnarray}
%   P(c_1 \mid f_{\mathrm{gas}}, D_L, \text{astrophysical inputs}) \nonumber  \\
%    \propto \mathcal{L}(\text{ $f_{gas}$, $D_L$, astrophysical inputs} \mid c_1) \times P_0(c_1),
%\end{eqnarray}

\begin{align}
P(c_1 \mid \mathbf{d})
\propto 
\mathcal{L}(\mathbf{d} \mid c_1)\, P_0(c_1),
\end{align}
where $\mathbf{d} \equiv \{ f_{\mathrm{gas}}, D_L, \text{astrophysical inputs} \}$,
where we adopt a flat prior in the range \(-1.2 \leq c_1 \leq 1.2\). In this framework, the inferred constraints on $c_1$ should be interpreted as conditional on the adopted astrophysical priors.} Although several astrophysical and calibration quantities enter the construction of $\xi$ (e.g., $K_{500}$, $\gamma$, $f_{\star,500}$, and $\Omega_b/\Omega_m$), in our implementation they are not sampled as additional MCMC dimensions. Instead, they are treated as external priors (see Table~\ref{priors}), and their uncertainties, including intrinsic scatters, are propagated into the total covariance $C_\xi$ through the Monte Carlo procedure used to build $\xi$ and $C_\xi$. Consequently, the \texttt{emcee} exploration is strictly one-dimensional, sampling only $c_1$. For this reason, a multi-parameter corner plot is not applicable here, and the relevant MCMC diagnostics reduce to the one-dimensional marginalized posterior of $c_1$. 

Our main results are summarized as follows:
\begin{itemize}
    \item {Mass bias $K_{500}^{CCCP}$:} {we obtain $c_1 = -0.10 \pm 0.12$ for ($P_1$) and $c_1 = -0.11 \pm 0.19$ for ($P_2$).} These results are illustrated in the left panel of Fig.~\ref{fig:results1}.

    \item {Mass bias $K_{500}^{CLASH}$:} {we obtain $c_1 = -0.03 \pm 0.12$ for ($P_1$) and $c_1 = 0.04 \pm 0.19$ for ($P_2$)}, as illustrated in the center panel of Fig.~\ref{fig:results1}.

    \item {Mass bias $K_{500}^{CMB}$:} {we obtain $c_1 = 0.20 \pm 0.13$ for ($P_1$) and $c_1 = 0.41 \pm 0.22$ for ($P_2$)}, as shown in the right panel of Fig.~\ref{fig:results1}.
\end{itemize}

These results, summarized in Table~\ref{tab:results}, illustrate how different assumptions regarding the mass bias parameter in the $f_{\mathrm{gas}}$ model influence constraints on a possible variation in the speed of light. When adopting the priors from the \textsc{CCCP} and \textsc{CLASH} analyses, we find that the parameter $c_1$ is consistent with zero within $1\sigma$, indicating no significant deviation from the standard constant value of $c$. In contrast, when using the prior $K_{500}^{\mathrm{CMB}}$, derived from a joint analysis of Planck primary CMB data, Planck-SZ number counts, the thermal SZ power spectrum, and BAO measurements~\cite{salvati2018constraints}, we observe a mild tension with the hypothesis of a constant speed of light, with consistency achieved only at the $2\sigma$ confidence level. This indicates that, in this case, the constancy of $c$ is only marginally supported by the data. Altogether, while some level of tension persists among the different mass-bias estimates, our results show that no statistically significant evidence for a variation in $c$ is found in any of the cases considered. These results underscore the pivotal role played by cluster mass bias in cosmological tests of fundamental physics and point to the need for improved, model-independent calibrations of this parameter in future surveys. {In Figs.~\ref{fig:delta_c_cccp_clash} and \ref{fig:delta_c_cmb} we show the evolution of $\Delta c/c$ for each case. As can be seen, for $K_{\rm CMB}$ the result is only marginally consistent with no variation at the $2\sigma$ confidence level.}

 We also tested the robustness of the current methodology by adopting a non-evolving value of $\gamma^{\text{The300}}_{500} = 0.938 \pm 0.041$, consistent with the analysis of simulated clusters from The Three Hundred Project \cite{cui2018three} as presented in \cite{eckert2019non}. In this case, for $K_{500}^{CCCP}$ we obtain $c_1 = -0.09 \pm 0.11$ and $c_1 = -0.11 \pm 0.19$ for $P_1$ and $P_2$, respectively. For CLASH, we find $c_1 = -0.01 \pm 0.12$ and $c_1 = 0.04 \pm 0.19$ for $P_1$ and $P_2$, respectively. Finally, for the CMB case, we obtain $c_1 = 0.24 \pm 0.13$ and $c_1 = 0.47 \pm 0.21$ for $P_1$ and $P_2$, respectively. These results are in good agreement with those derived using $\gamma^{\text{FABLE}}_{500}(z)$.

\subsubsection{Obtaing $K_{X-COP}$ by considering a $c$ constant }

 {As a complementary consistency test, we assume a constant speed of light in Eq.~\eqref{fgasphi} and infer the mass-bias parameter required by the X-COP sample. Under this assumption, we obtain $K_{\text{X-COP}} = 0.96 \pm 0.17$ ($1\sigma$). Although the associated uncertainty is relatively large, this result is consistent with previous mass-calibration estimates for the X-COP sample within the $1\sigma$ confidence level \cite{eckert2019non}. This finding indicates that, in the absence of variations in $c$, the X-COP data do not favor a significant departure from the standard cluster mass calibration.}

\begin{table}[]
    \centering
    \begin{tabular}{c c}
         Parameter & Prior \\
         \hline
         \hline
         \rule{0pt}{3ex}
          $c_1$ & $\mathcal{U}[-1.200;1.200]$ \\
         $\Omega_{b}/\Omega_{m}$ & $\mathcal{N}[0.173;0.027]$ \\
         $\gamma_0$ & $\mathcal{N}[0.931;0.040]$ \\
         $K^{CCCP}_{500}$ & $\mathcal{N}[0.840;0.040]$ \\
         $K^{CLASH}_{500}$ & $\mathcal{N}[0.780;0.090]$ \\
         $K^{CMB}_{500}$ & $\mathcal{N}[0.650;0.040]$ \\
         $f_{\star,500}$ & $\mathcal{N}[0.015;0.005]$ \\
         \hline
    \end{tabular}
   \caption{ Priors adopted for external inputs entering the construction of $\xi$ and its covariance $C_\xi$. Gaussian priors are denoted by $\mathcal{N}$[mean; standard deviation], while flat priors are denoted by $\mathcal{U}$[min; max].  Only $c_1$ is sampled in the MCMC; the remaining quantities are treated as external priors and are marginalized through the Monte Carlo propagation into $C_\xi$ (see Sec.~\ref{sec:analysis}).}
    \label{priors}
\end{table}

\section{Conclusions}
\label{sec:conclusions}

In this work, we investigated the constancy of the speed of light by combining gas mass fraction measurements from galaxy clusters with {SNe Ia data from the Pantheon+ sample} \cite{Brout:2022vxf}.  The galaxy clusters data consists of gas mass fraction measurements of 12 clusters at $z \lesssim 0.1$ from X-COP \cite{eckert2019non}, 44 clusters in the range $0.1 \lesssim z \lesssim 0.3$ from \cite{ettori2010mass}, and 47 clusters at $0.4 \lesssim z \lesssim 1.2$ from \cite{ghirardini2017evolution}.

We explored two phenomenological parameterizations for a possible redshift dependence of the speed of light, \(c(z) = c_0(1 + c_1 z)\) and \(c(z) = c_0(1 + c_1 z / (1 + z))\), assessing how different Gaussian priors on the cluster mass bias—namely, \(K_{500}^{\mathrm{CLASH}}\), \(K_{500}^{\mathrm{CCCP}}\), and \(K_{500}^{\mathrm{CMB}}\)—affect the inferred constraints. Our analysis shows that no significant deviation from a constant speed of light is found when adopting the \textsc{CLASH} or \textsc{CCCP} priors. When using the prior \(K_{500}^{\mathrm{CMB}}\), derived from Planck-based analyses, the results mildly disfavor the hypothesis of a constant \(c\), with consistency achieved only at the \(2\sigma\) confidence level. This highlights the pivotal role of mass bias in cosmological tests of fundamental physics.

These findings reinforce the importance of accurate mass calibration in cluster cosmology and underscore the need to reduce systematic uncertainties through independent and complementary observational strategies. Upcoming surveys---including high-resolution CMB lensing, deep X-ray observations, and wide-field gravitational lensing campaigns---will be crucial for refining cluster mass estimates and enabling robust tests of fundamental principles such as the invariance of the speed of light. {With eROSITA~\cite{2012arXiv1209.3114M} already carrying out an all-sky X-ray survey of thousands of galaxy clusters, increasingly precise gas-mass-fraction measurements are becoming available, which will render the analysis proposed here more robust.} Future progress in this direction will depend less on further phenomenological parametrizations of \(c(z)\) and more on improving the control of cluster astrophysics, sample homogeneity, and mass calibration systematics.
\section*{ACKNOWLEDGEMENT}
RFLH thanks Conselho Nacional de Desenvolvimento Cient\'ifico e Tecnol\'ogico (CNPq), No.428755/2018-6 and 305930/2017-6. SHP also acknowledges financial support from CNPq, No. 308469/2021-6.

\bibliography{ref}

@misc{dinda2024modelagnosticassessmentdarkenergy,
      title={Model-agnostic assessment of dark energy after DESI DR1 BAO}, 
      author={Bikash R. Dinda and Roy Maartens},
      year={2024},
      eprint={2407.17252},
      archivePrefix={arXiv},
      primaryClass={astro-ph.CO},
      url={https://arxiv.org/abs/2407.17252}, 
}

@article{eckert2016xxl,
  title={The XXL Survey-XIII. Baryon content of the bright cluster sample},
  author={Eckert, D and Ettori, STEFANO and Coupon, J and Gastaldello, FABIO and Pierre, M and Melin, J-B and Le Brun, AMC and McCarthy, IG and Adami, C and Chiappetti, LUCIO and others},
  journal={Astronomy \& Astrophysics},
  volume={592},
  pages={A12},
  year={2016},
  publisher={EDP Sciences}
}

@article{Chiu_2018,
   title={Baryon content in a sample of 91 galaxy clusters selected by the South Pole Telescope at 0.2 &lt;z &lt; 1.25},
   volume={478},
   ISSN={1365-2966},
   url={http://dx.doi.org/10.1093/mnras/sty1284},
   DOI={10.1093/mnras/sty1284},
   number={3},
   journal={Monthly Notices of the Royal Astronomical Society},
   publisher={Oxford University Press (OUP)},
   author={Chiu, I and Mohr, J J and McDonald, M and Bocquet, S and Desai, S and Klein, M and Israel, H and Ashby, M L N and Stanford, A and Benson, B A and Brodwin, M and Abbott, T M C and Abdalla, F B and Allam, S and Annis, J and Bayliss, M and Benoit-Lévy, A and Bertin, E and Bleem, L and Brooks, D and Buckley-Geer, E and Bulbul, E and Capasso, R and Carlstrom, J E and Rosell, A Carnero and Carretero, J and Castander, F J and Cunha, C E and D’Andrea, C B and da Costa, L N and Davis, C and Diehl, H T and Dietrich, J P and Doel, P and Drlica-Wagner, A and Eifler, T F and Evrard, A E and Flaugher, B and García-Bellido, J and Garmire, G and Gaztanaga, E and Gerdes, D W and Gonzalez, A and Gruen, D and Gruendl, R A and Gschwend, J and Gupta, N and Gutierrez, G and Hlavacek-L, J and Honscheid, K and James, D J and Jeltema, T and Kraft, R and Krause, E and Kuehn, K and Kuhlmann, S and Kuropatkin, N and Lahav, O and Lima, M and Maia, M A G and Marshall, J L and Melchior, P and Menanteau, F and Miquel, R and Murray, S and Nord, B and Ogando, R L C and Plazas, A A and Rapetti, D and Reichardt, C L and Romer, A K and Roodman, A and Sanchez, E and Saro, A and Scarpine, V and Schindler, R and Schubnell, M and Sharon, K and Smith, R C and Smith, M and Soares-Santos, M and Sobreira, F and Stalder, B and Stern, C and Strazzullo, V and Suchyta, E and Swanson, M E C and Tarle, G and Vikram, V and Walker, A R and Weller, J and Zhang, Y},
   year={2018},
   month=may, pages={3072–3099} }

@article{Leauthaud_2011,
   title={NEW CONSTRAINTS ON THE EVOLUTION OF THE STELLAR-TO-DARK MATTER CONNECTION: A COMBINED ANALYSIS OF GALAXY-GALAXY LENSING, CLUSTERING, AND STELLAR MASS FUNCTIONS FROMz= 0.2 toz= 1},
   volume={744},
   ISSN={1538-4357},
   url={http://dx.doi.org/10.1088/0004-637X/744/2/159},
   DOI={10.1088/0004-637x/744/2/159},
   number={2},
   journal={The Astrophysical Journal},
   publisher={American Astronomical Society},
   author={Leauthaud, Alexie and Tinker, Jeremy and Bundy, Kevin and Behroozi, Peter S. and Massey, Richard and Rhodes, Jason and George, Matthew R. and Kneib, Jean-Paul and Benson, Andrew and Wechsler, Risa H. and Busha, Michael T. and Capak, Peter and Cortês, Marina and Ilbert, Olivier and Koekemoer, Anton M. and Le Fèvre, Oliver and Lilly, Simon and McCracken, Henry J. and Salvato, Mara and Schrabback, Tim and Scoville, Nick and Smith, Tristan and Taylor, James E.},
   year={2011},
   month=dec, pages={159} }

@article{Gonzalez:2024qjs,
    author = "Gonzalez, Javier E. and Ferreira, Marcelo and Cola\c{c}o, Leorando R. and Holanda, Rodrigo F. L. and Nunes, Rafael C.",
    title = "{Unveiling the Hubble constant through galaxy cluster gas mass fractions}",
    eprint = "2405.13665",
    archivePrefix = "arXiv",
    primaryClass = "astro-ph.CO",
    doi = "10.1016/j.physletb.2024.138982",
    journal = "Phys. Lett. B",
    volume = "857",
    pages = "138982",
    year = "2024"
}

@article{riess2022,
  title={A comprehensive measurement of the local value of the Hubble constant with 1 km s- 1 Mpc- 1 uncertainty from the Hubble Space Telescope and the SH0ES team},
  author={Riess, Adam G and Yuan, Wenlong and Macri, Lucas M and Scolnic, Dan and Brout, Dillon and Casertano, Stefano and Jones, David O and Murakami, Yukei and Anand, Gagandeep S and Breuval, Louise and others},
  journal={The Astrophysical journal letters},
  volume={934},
  number={1},
  pages={L7},
  year={2022},
  publisher={IOP Publishing}
}

@article{Brout:2022vxf,
    author = "Brout, Dillon and others",
    title = "{The Pantheon+ Analysis: Cosmological Constraints}",
    eprint = "2202.04077",
    archivePrefix = "arXiv",
    primaryClass = "astro-ph.CO",
    month = "2",
    year = "2022",
    journal = ""
}

@ARTICLE{2018A&A...610L...4H,
       author = {{Hurier}, G. and {Angulo}, R.~E.},
        title = "{Measuring the hydrostatic mass bias in galaxy clusters by combining Sunyaev-Zel'dovich and CMB lensing data}",
      journal = {\aap},
         year = 2018,
        month = feb,
       volume = {610},
          eid = {L4},
        pages = {L4},
          doi = {10.1051/0004-6361/201731999},
archivePrefix = {arXiv},
       eprint = {1711.06029},
 primaryClass = {astro-ph.CO},
       adsurl = {https://ui.adsabs.harvard.edu/abs/2018A&A...610L...4H}
}

@ARTICLE{2015ApJ...806..247B,
       author = {{Baxter}, E.~J. and {Keisler}, R. and {Dodelson}, S. and {Aird}, K.~A. and {Allen}, S.~W. and {Ashby}, M.~L.~N. and {Bautz}, M. and {Bayliss}, M. and {Benson}, B.~A. and {Bleem}, L.~E. and {Bocquet}, S. and {Brodwin}, M. and {Carlstrom}, J.~E. and {Chang}, C.~L. and {Chiu}, I. and {Cho}, H. -M. and {Clocchiatti}, A. and {Crawford}, T.~M. and {Crites}, A.~T. and {Desai}, S. and {Dietrich}, J.~P. and {de Haan}, T. and {Dobbs}, M.~A. and {Foley}, R.~J. and {Forman}, W.~R. and {George}, E.~M. and {Gladders}, M.~D. and {Gonzalez}, A.~H. and {Halverson}, N.~W. and {Harrington}, N.~L. and {Hennig}, C. and {Hoekstra}, H. and {Holder}, G.~P. and {Holzapfel}, W.~L. and {Hou}, Z. and {Hrubes}, J.~D. and {Jones}, C. and {Knox}, L. and {Lee}, A.~T. and {Leitch}, E.~M. and {Liu}, J. and {Lueker}, M. and {Luong-Van}, D. and {Mantz}, A. and {Marrone}, D.~P. and {McDonald}, M. and {McMahon}, J.~J. and {Meyer}, S.~S. and {Millea}, M. and {Mocanu}, L.~M. and {Murray}, S.~S. and {Padin}, S. and {Pryke}, C. and {Reichardt}, C.~L. and {Rest}, A. and {Ruhl}, J.~E. and {Saliwanchik}, B.~R. and {Saro}, A. and {Sayre}, J.~T. and {Schaffer}, K.~K. and {Shirokoff}, E. and {Song}, J. and {Spieler}, H.~G. and {Stalder}, B. and {Stanford}, S.~A. and {Staniszewski}, Z. and {Stark}, A.~A. and {Story}, K.~T. and {van Engelen}, A. and {Vanderlinde}, K. and {Vieira}, J.~D. and {Vikhlinin}, A. and {Williamson}, R. and {Zahn}, O. and {Zenteno}, A.},
        title = "{A Measurement of Gravitational Lensing of the Cosmic Microwave Background by Galaxy Clusters Using Data from the South Pole Telescope}",
      journal = {\apj},
         year = 2015,
        month = jun,
       volume = {806},
       number = {2},
          eid = {247},
        pages = {247},
          doi = {10.1088/0004-637X/806/2/247},
archivePrefix = {arXiv},
       eprint = {1412.7521},
 primaryClass = {astro-ph.CO},
       adsurl = {https://ui.adsabs.harvard.edu/abs/2015ApJ...806..247B}
}

@ARTICLE{2019MNRAS.489..401Z,
       author = {{Zubeldia}, {\'I}{\~n}igo and {Challinor}, Anthony},
        title = "{Cosmological constraints from Planck galaxy clusters with CMB lensing mass bias calibration}",
      journal = {\mnras},
         year = 2019,
        month = oct,
       volume = {489},
       number = {1},
        pages = {401-419},
          doi = {10.1093/mnras/stz2153},
archivePrefix = {arXiv},
       eprint = {1904.07887},
 primaryClass = {astro-ph.CO},
       adsurl = {https://ui.adsabs.harvard.edu/abs/2019MNRAS.489..401Z}
}

@ARTICLE{2023A&A...674A..48W,
       author = {{Wicker}, R. and {Douspis}, M. and {Salvati}, L. and {Aghanim}, N.},
        title = "{Constraining the mass and redshift evolution of the hydrostatic mass bias using the gas mass fraction in galaxy clusters}",
      journal = {\aap},
         year = 2023,
        month = jun,
       volume = {674},
          eid = {A48},
        pages = {A48},
          doi = {10.1051/0004-6361/202243922},
archivePrefix = {arXiv},
       eprint = {2204.12823},
 primaryClass = {astro-ph.CO},
       adsurl = {https://ui.adsabs.harvard.edu/abs/2023A&A...674A..48W}
}

@article{10.1093/mnras/staa2235,
    author = {Henden, Nicholas A and Puchwein, Ewald and Sijacki, Debora},
    title = {The baryon content of groups and clusters of galaxies in the FABLE simulations},
    journal = {Monthly Notices of the Royal Astronomical Society},
    volume = {498},
    number = {2},
    pages = {2114-2137},
    year = {2020},
    month = {09},
    issn = {0035-8711},
    doi = {10.1093/mnras/staa2235},
    url = {https://doi.org/10.1093/mnras/staa2235},
    eprint = {https://academic.oup.com/mnras/article-pdf/498/2/2114/33776942/staa2235.pdf},
}

@article{cui2018three,
  title={The Three Hundred project},
  author={Cui, Weiguang and Knebe, Alexander and Yepes, Gustavo and Pearce, Frazer and Power, Chris and Dave, Romeel and Arth, Alexander and Borgani, Stefano and Dolag, Klaus and Elahi, Pascal and others},
  journal={Monthly Notices of the Royal Astronomical Society},
  volume={480},
  year={2018}
}

@article{eckert2019non,
  title={Non-thermal pressure support in X-COP galaxy clusters},
  author={Eckert, D and Ghirardini, V and Ettori, Stefano and Rasia, Elena and Biffi, V and Pointecouteau, E and Rossetti, Mariachiara and Molendi, Silvano and Vazza, Franco and Gastaldello, Fabio and others},
  journal={Astronomy \& Astrophysics},
  volume={621},
  pages={A40},
  year={2019},
  publisher={EDP Sciences}
}

@ARTICLE{2008MNRAS.383..879A,
       author = {{Allen}, S.~W. and {Rapetti}, D.~A. and {Schmidt}, R.~W. and {Ebeling}, H. and {Morris}, R.~G. and {Fabian}, A.~C.},
        title = "{Improved constraints on dark energy from Chandra X-ray observations of the largest relaxed galaxy clusters}",
      journal = {\mnras},
         year = 2008,
        month = jan,
       volume = {383},
       number = {3},
        pages = {879-896},
          doi = {10.1111/j.1365-2966.2007.12610.x},
archivePrefix = {arXiv},
       eprint = {0706.0033},
 primaryClass = {astro-ph},
       adsurl = {https://ui.adsabs.harvard.edu/abs/2008MNRAS.383..879A}
}

@article{corasaniti2021cosmological,
  title={Cosmological constraints from galaxy cluster sparsity, cluster gas mass fraction, and baryon acoustic oscillation data},
  author={Corasaniti, Pier-Stefano and Sereno, Mauro and Ettori, Stefano},
  journal={The Astrophysical Journal},
  volume={911},
  number={2},
  pages={82},
  year={2021},
  publisher={IOP Publishing}
}

@article{sereno2015comparing,
  title={Comparing masses in literature (CoMaLit)--I. Bias and scatter in weak lensing and X-ray mass estimates of clusters},
  author={Sereno, Mauro and Ettori, Stefano},
  journal={Monthly Notices of the Royal Astronomical Society},
  volume={450},
  number={4},
  pages={3633--3648},
  year={2015},
  publisher={Oxford University Press}
}

@article{salvati2018constraints,
  title={Constraints from thermal Sunyaev-Zel’dovich cluster counts and power spectrum combined with CMB},
  author={Salvati, Laura and Douspis, Marian and Aghanim, Nabila},
  journal={Astronomy \& Astrophysics},
  volume={614},
  pages={A13},
  year={2018},
  publisher={EDP Sciences}
}

@article{herbonnet2020cccp,
  title={CCCP and MENeaCS:(updated) weak-lensing masses for 100 galaxy clusters},
  author={Herbonnet, Ricardo and Sif{\'o}n, Crist{\'o}bal and Hoekstra, Henk and Bah{\'e}, Yannick and van Der Burg, Remco FJ and Melin, Jean-Baptiste and von Der Linden, Anja and Sand, David and Kay, Scott and Barnes, David},
  journal={Monthly Notices of the Royal Astronomical Society},
  volume={497},
  number={4},
  pages={4684--4703},
  year={2020},
  publisher={Oxford University Press}
}

@article{ettori2010mass,
  title={Mass profiles and c- MDM relation in X-ray luminous galaxy clusters},
  author={Ettori, S and Gastaldello, F and Leccardi, A and Molendi, S and Rossetti, M and Buote, D and Meneghetti, M},
  journal={Astronomy \& Astrophysics},
  volume={524},
  pages={A68},
  year={2010},
  publisher={EDP Sciences}
}

@article{ghirardini2017evolution,
  title={On the evolution of the entropy and pressure profiles in X-ray luminous galaxy clusters at z> 0.4},
  author={Ghirardini, V and Ettori, STEFANO and Amodeo, S and Capasso, RAFFAELLA and Sereno, Mauro},
  journal={Astronomy \& Astrophysics},
  volume={604},
  pages={A100},
  year={2017},
  publisher={Edp Sciences}
}

@article{foreman2013emcee,
  title={emcee: the MCMC hammer},
  author={Foreman-Mackey, Daniel and Hogg, David W and Lang, Dustin and Goodman, Jonathan},
  journal={Publications of the Astronomical Society of the Pacific},
  volume={125},
  number={925},
  pages={306},
  year={2013},
  publisher={IOP Publishing}
}

@article{seikel2012reconstruction,
  title={Reconstruction of dark energy and expansion dynamics using Gaussian processes},
  author={Seikel, Marina and Clarkson, Chris and Smith, Mathew},
  journal={Journal of Cosmology and Astroparticle Physics},
  volume={2012},
  number={06},
  pages={036},
  year={2012},
  publisher={IOP Publishing}
}

@article{sarazin1986x,
  title={X-ray emission from clusters of galaxies},
  author={Sarazin, Craig L},
  journal={Reviews of Modern Physics},
  volume={58},
  number={1},
  pages={1},
  year={1986},
  publisher={APS}
}

@ARTICLE{2011ARA&A..49..409A,
       author = {{Allen}, Steven W. and {Evrard}, August E. and {Mantz}, Adam B.},
        title = "{Cosmological Parameters from Observations of Galaxy Clusters}",
      journal = {\araa},
         year = 2011,
        month = sep,
       volume = {49},
       number = {1},
        pages = {409-470},
          doi = {10.1146/annurev-astro-081710-102514},
archivePrefix = {arXiv},
       eprint = {1103.4829},
 primaryClass = {astro-ph.CO},
       adsurl = {https://ui.adsabs.harvard.edu/abs/2011ARA&A..49..409A}
}

@article{Dzuba:1998au,
    author = "Dzuba, V. A. and Flambaum, V. V. and Webb, J. K.",
    title = "{Calculations of the relativistic effects in many electron atoms and space-time variation of fundamental constants}",
    eprint = "physics/9808021",
    archivePrefix = "arXiv",
    doi = "10.1103/PhysRevA.59.230",
    journal = "Phys. Rev. A",
    volume = "59",
    pages = "230--237",
    year = "1999"
}

@article{Webb:1998cq,
    author = "Webb, John K. and Flambaum, Victor V. and Churchill, Christopher W. and Drinkwater, Michael J. and Barrow, John D.",
    title = "{Evidence for time variation of the fine structure constant}",
    eprint = "astro-ph/9803165",
    archivePrefix = "arXiv",
    reportNumber = "AST-140398",
    doi = "10.1103/PhysRevLett.82.884",
    journal = "Phys. Rev. Lett.",
    volume = "82",
    pages = "884--887",
    year = "1999"
}

@article{Liu:2021eit,
    author = "Liu, Tonghua and Cao, Shuo and Biesiada, Marek and Liu, Yuting and Lian, Yujie and Zhang, Yilong",
    title = "{Consistency testing for invariance of the speed of light at different redshifts: the newest results from strong lensing and Type Ia supernovae observations}",
    eprint = "2106.15145",
    archivePrefix = "arXiv",
    primaryClass = "astro-ph.CO",
    doi = "10.1093/mnras/stab1868",
    journal = "Mon. Not. Roy. Astron. Soc.",
    volume = "506",
    number = "2",
    pages = "2181--2188",
    year = "2021"
}

@article{Mendonca:2021eux,
    author = "Mendon\c{c}a, I. E. C. R. and Bora, Kamal and Holanda, R. F. L. and Desai, Shantanu and Pereira, S. H.",
    title = "{A search for the variation of speed of light using galaxy cluster gas mass fraction measurements}",
    eprint = "2109.14512",
    archivePrefix = "arXiv",
    primaryClass = "astro-ph.CO",
    doi = "10.1088/1475-7516/2021/11/034",
    journal = "JCAP",
    volume = "11",
    pages = "034",
    year = "2021"
}

@ARTICLE{2017RPPh...80l6902M,
       author = {{Martins}, C.~J.~A.~P.},
        title = "{The status of varying constants: a review of the physics, searches and implications}",
      journal = {Reports on Progress in Physics},
         year = 2017,
        month = dec,
       volume = {80},
       number = {12},
          eid = {126902},
        pages = {126902},
          doi = {10.1088/1361-6633/aa860e},
archivePrefix = {arXiv},
       eprint = {1709.02923},
 primaryClass = {astro-ph.CO},
       adsurl = {https://ui.adsabs.harvard.edu/abs/2017RPPh...80l6902M}
}

@article{Uzan:2010pm,
    author = "Uzan, Jean-Philippe",
    title = "{Varying Constants, Gravitation and Cosmology}",
    eprint = "1009.5514",
    archivePrefix = "arXiv",
    primaryClass = "astro-ph.CO",
    doi = "10.12942/lrr-2011-2",
    journal = "Living Rev. Rel.",
    volume = "14",
    pages = "2",
    year = "2011"
}

@article{Rodrigues:2021wyk,
    author = "Rodrigues, Gabriel and Bengaly, Carlos",
    title = "{A model-independent test of speed of light variability with cosmological observations}",
    eprint = "2112.01963",
    archivePrefix = "arXiv",
    primaryClass = "astro-ph.CO",
    doi = "10.1088/1475-7516/2022/07/029",
    journal = "JCAP",
    volume = "07",
    number = "07",
    pages = "029",
    year = "2022"
}

@article{Allen2007,
 
  year = {2007},
  month = dec,
  publisher = {Oxford University Press ({OUP})},
  volume = {383},
  number = {3},
  pages = {879--896},
  author = {S. W. Allen and D. A. Rapetti and R. W. Schmidt and H. Ebeling and R. G. Morris and A. C. Fabian},
  title = {Improved constraints on dark energy from Chandra X-ray observations of the largest relaxed galaxy clusters},
  journal = {Monthly Notices of the Royal Astronomical Society}
}

@ARTICLE{Ettori2009,
       author = {{Ettori}, S. and {Morandi}, A. and {Tozzi}, P. and {Balestra}, I. and {Borgani}, S. and {Rosati}, P. and {Lovisari}, L. and {Terenziani}, F.},
        title = "{The cluster gas mass fraction as a cosmological probe: a revised study}",
      journal = {\aap},
         year = 2009,
        month = jul,
       volume = {501},
       number = {1},
        pages = {61-73},
          doi = {10.1051/0004-6361/200810878},
archivePrefix = {arXiv},
       eprint = {0904.2740},
 primaryClass = {astro-ph.CO},
       adsurl = {https://ui.adsabs.harvard.edu/abs/2009A&A...501...61E}
}

@ARTICLE{Allen2011,
       author = {{Allen}, Steven W. and {Evrard}, August E. and {Mantz}, Adam B.},
        title = "{Cosmological Parameters from Observations of Galaxy Clusters}",
      journal = {\araa},
         year = 2011,
        month = sep,
       volume = {49},
       number = {1},
        pages = {409-470},
          doi = {10.1146/annurev-astro-081710-102514},
archivePrefix = {arXiv},
       eprint = {1103.4829},
 primaryClass = {astro-ph.CO},
       adsurl = {https://ui.adsabs.harvard.edu/abs/2011ARA&A..49..409A}
}

@ARTICLE{2012arXiv1209.3114M,
       author = {{Merloni}, A. and {Predehl}, P. and {Becker}, W. and {B{\"o}hringer}, H. and {Boller}, T. and {Brunner}, H. and {Brusa}, M. and {Dennerl}, K. and {Freyberg}, M. and {Friedrich}, P. and {Georgakakis}, A. and {Haberl}, F. and {Hasinger}, G. and {Meidinger}, N. and {Mohr}, J. and {Nandra}, K. and {Rau}, A. and {Reiprich}, T.~H. and {Robrade}, J. and {Salvato}, M. and {Santangelo}, A. and {Sasaki}, M. and {Schwope}, A. and {Wilms}, J. and {German eROSITA Consortium}, the},
        title = "{eROSITA Science Book: Mapping the Structure of the Energetic Universe}",
      journal = {arXiv e-prints},
         year = 2012,
        month = sep,
          eid = {arXiv:1209.3114},
        pages = {arXiv:1209.3114},
          doi = {10.48550/arXiv.1209.3114},
archivePrefix = {arXiv},
       eprint = {1209.3114},
 primaryClass = {astro-ph.HE},
       adsurl = {https://ui.adsabs.harvard.edu/abs/2012arXiv1209.3114M}
}

@ARTICLE{2025PhRvD.111f3526K,
       author = {{Krolewski}, Alex and {Percival}, Will J.},
        title = "{Measuring the baryon fraction using galaxy clustering}",
      journal = {\prd},
         year = 2025,
        month = mar,
       volume = {111},
       number = {6},
          eid = {063526},
        pages = {063526},
          doi = {10.1103/PhysRevD.111.063526},
archivePrefix = {arXiv},
       eprint = {2403.19236},
 primaryClass = {astro-ph.CO},
       adsurl = {https://ui.adsabs.harvard.edu/abs/2025PhRvD.111f3526K}
}

@ARTICLE{2022A&A...663L...6A,
       author = {{Angelinelli}, M. and {Ettori}, S. and {Dolag}, K. and {Vazza}, F. and {Ragagnin}, A.},
        title = "{Mapping `out-of-the-box' the properties of the baryons in massive halos}",
      journal = {\aap},
         year = 2022,
        month = jul,
       volume = {663},
          eid = {L6},
        pages = {L6},
          doi = {10.1051/0004-6361/202244068},
archivePrefix = {arXiv},
       eprint = {2206.08382},
 primaryClass = {astro-ph.GA},
       adsurl = {https://ui.adsabs.harvard.edu/abs/2022A&A...663L...6A}
}

@article{Mantz:2014xba,
    author = "Mantz, A. B. and Allen, S. W. and Morris, R. G. and Rapetti, D. A. and Applegate, D. E. and Kelly, P. L. and von der Linden, Anja and Schmidt, R. W.",
    title = "{Cosmology and astrophysics from relaxed galaxy clusters \textendash{} II. Cosmological constraints}",
    eprint = "1402.6212",
    archivePrefix = "arXiv",
    primaryClass = "astro-ph.CO",
    doi = "10.1093/mnras/stu368",
    journal = "Mon. Not. Roy. Astron. Soc.",
    volume = "440",
    number = "3",
    pages = "2077--2098",
    year = "2014"
}

@article{King:2012id,
    author = "King, Julian A. and Webb, John K. and Murphy, Michael T. and Flambaum, Victor V. and Carswell, Robert F. and Bainbridge, Matthew B. and Wilczynska, Michael R. and Koch, F. Elliot",
    title = "{Spatial variation in the fine-structure constant -- new results from VLT/UVES}",
    eprint = "1202.4758",
    archivePrefix = "arXiv",
    primaryClass = "astro-ph.CO",
    doi = "10.1111/j.1365-2966.2012.20852.x",
    journal = "Mon. Not. Roy. Astron. Soc.",
    volume = "422",
    pages = "3370--3413",
    year = "2012"
}

@article{Kotus:2016xxb,
    author = "Kotu\v{s}, Sr\dj{}an M. and Murphy, Michael T. and Carswell, Robert F.",
    title = "{High-precision limit on variation in the fine-structure constant from a single quasar absorption system}",
    eprint = "1609.03860",
    archivePrefix = "arXiv",
    primaryClass = "astro-ph.CO",
    doi = "10.1093/mnras/stw2543",
    journal = "Mon. Not. Roy. Astron. Soc.",
    volume = "464",
    number = "3",
    pages = "3679--3703",
    year = "2017"
}

@article{Salzano:2016hce,
    author = "Salzano, Vincenzo",
    title = "{Recovering a redshift-extended varying speed of light signal from galaxy surveys}",
    eprint = "1604.03398",
    archivePrefix = "arXiv",
    primaryClass = "astro-ph.CO",
    doi = "10.1103/PhysRevD.95.084035",
    journal = "Phys. Rev. D",
    volume = "95",
    number = "8",
    pages = "084035",
    year = "2017"
}

@article{Qi:2014zja,
    author = "Qi, Jing-Zhao and Zhang, Ming-Jian and Liu, Wen-Biao",
    title = "{Observational constraint on the varying speed of light theory}",
    eprint = "1407.1265",
    archivePrefix = "arXiv",
    primaryClass = "gr-qc",
    doi = "10.1103/PhysRevD.90.063526",
    journal = "Phys. Rev. D",
    volume = "90",
    number = "6",
    pages = "063526",
    year = "2014"
}

@article{Wang:2019tdn,
    author = "Wang, Dandan and Zhang, Hanyu and Zheng, Jinglan and Wang, Yuting and Zhao, Gong-Bo",
    title = "{A model independent constraint on the temporal evolution of the speed of light}",
    eprint = "1904.04041",
    archivePrefix = "arXiv",
    primaryClass = "astro-ph.CO",
    month = "4",
    year = "2019"
}

@article{Galli:2012bf,
    author = "Galli, Silvia",
    title = "{Clusters of galaxies and variation of the fine structure constant}",
    eprint = "1212.1075",
    archivePrefix = "arXiv",
    primaryClass = "astro-ph.CO",
    doi = "10.1103/PhysRevD.87.123516",
    journal = "Phys. Rev. D",
    volume = "87",
    number = "12",
    pages = "123516",
    year = "2013"
}

@ARTICLE{emcee,
       author = {{Foreman-Mackey}, Daniel and {Hogg}, David W. and {Lang}, Dustin and
         {Goodman}, Jonathan},
        title = "{emcee: The MCMC Hammer}",
      journal = {\pasp},
         year = 2013,
        month = mar,
       volume = {125},
       number = {925},
        pages = {306},
          doi = {10.1086/670067},
archivePrefix = {arXiv},
       eprint = {1202.3665},
 primaryClass = {astro-ph.IM},
       adsurl = {https://ui.adsabs.harvard.edu/abs/2013PASP..125..306F}
}

\end{document}